# Interfacing CRYSTAL/AMBER to Optimize QM/MM Lennard–Jones Parameters for Water and to Study Solvation of TiO$_2$ Nanoparticles

Asmus Ougaard Dohn [1,*], Daniele Selli [2], Gianluca Fazio [2], Lorenzo Ferraro [2], Jens Jørgen Mortensen [3], Bartolomeo Civalleri [4] and Cristiana Di Valentin [2]

1   Faculty of Physical Sciences and Science Institute, University of Iceland, 107 Reykjavík, Iceland
2   Dipartimento di Scienza dei Materiali, Università di Milano-Bicocca, via Cozzi 55, 20125 Milano, Italy; daniele.selli@unimib.it (D.S.); g.fazio3@campus.unimib.it (G.F.); lorenzo.ferraro@unimib.it (L.F.); cristiana.divalentin@unimib.it (C.D.V.)
3   CAMD, Department of Physics, Technical University of Denmark, 2800 Kongens Lyngby, Denmark; jjmo@dtu.dk
4   Dipartimento di Chimica, Università di Torino and NIS Centre of Excellence, Via P. Giuria 7, I-10129 Torino, Italy; bartolomeo.civalleri@unito.it
*   Correspondence: asod@hi.is; Tel.: +354-8379337

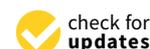

**Abstract:** Metal oxide nanoparticles (NPs) are regarded as good candidates for many technological applications, where their functional environment is often an aqueous solution. The correct description of metal oxide electronic structure is still a challenge for local and semilocal density functionals, whereas hybrid functional methods provide an improved description, and local atomic function-based codes such as CRYSTAL17 outperform plane wave codes when it comes to hybrid functional calculations. However, the computational cost of hybrids are still prohibitive for systems of real sizes, in a real environment. Therefore, we here present and critically assess the accuracy of our electrostatic embedding quantum mechanical/molecular mechanical (QM/MM) coupling between CRYSTAL17 and AMBER16, and demonstrate some of its capabilities via the case study of TiO$_2$ NPs in water. First, we produced new Lennard–Jones (LJ) parameters that improve the accuracy of water–water interactions in the B3LYP/TIP3P coupling. We found that optimizing LJ parameters based on water tri- to deca-mer clusters provides a less overstructured QM/MM liquid water description than when fitting LJ parameters only based on the water dimer. Then, we applied our QM/MM coupling methodology to describe the interaction of a 1 nm wide multilayer of water surrounding a spherical TiO$_2$ nanoparticle (NP). Optimizing the QM/MM water–water parameters was found to have little to no effect on the local NP properties, which provide insights into the range of influence that can be attributed to the LJ term in the QM/MM coupling. The effect of adding additional water in an MM fashion on the geometry optimized nanoparticle structure is small, but more evident effects are seen in its electronic properties. We also show that there is good transferability of existing QM/MM LJ parameters for organic molecules–water interactions to our QM/MM implementation, even though these parameters were obtained with a different QM code and QM/MM implementation, but with the same functional.

**Keywords:** QM/MM; multiscale; nanoparticles; force field parameters; water; titanium dioxide; geometry optimization; molecular dynamics

## 1. Introduction

After the pioneering work of M. Levitt and A. Warshel [1], the fundamental QM/MM methodology of joining explicit electronic structure to classical potential functions is constantly being both applied and expanded upon [2–7]. While much work is being put into methods that go beyond the electrostatic embedding-type of additive QM/MM, which explicitly couples the two subsystems via a Coloumb term but leaves the MM charges fixed, this form of coupling still seems prevalent within additive QM/MM applications [8], most likely due to its generally accepted trade-off between accuracy and efficiency. Some of the authors involved in this work have previously demonstrated how electrostatic embedding QM/MM can be used to predict solvation dynamics of bond formation in a model catalyst system, which was later confirmed by experiment [9,10]. However, the QM/MM implementation employed previously [11] relies on a DFT code that at the time of writing does not yet support the calculation of forces from hybrid functionals. In contrast, the MPP-CRYSTAL17 [12–14] code has an efficient implementation of the exact exchange term [12]. This capability is crucial for successfully modeling semiconducting oxide systems, such as $TiO_2$ nanoparticles (NPs) in aqueous solution [15]. Metal oxide nanoparticles NPs are widely used in many technological applications, and especially $TiO_2$ nanomaterials are very popular because of their abundance, nontoxicity, high stability under a variety of conditions, and biocompatibility. Their uses ranges from traditional ones, such as in the pigment industry or in cosmetics, to more advanced fields, such as photocatalysis, photoelectrochemistry, fuel production, and nanomedicine [16–19]. Many, if not all, of the applications take place in aqueous environments, where water is often a main actor in the physics and chemistry of such processes [20,21]. However, the modeling of $TiO_2$ nanoparticles is computationally very demanding [22–25]: apart from requiring hybrid functionals, realistic models consist of 500–1000 atoms for NPs of at least 2 nm of diameter, which cannot be treated periodically. Therefore, such simulations require the use of parallelized, efficiently performing codes, such as MPP-CRYSTAL17. However, when also including the aqueous environment of the NP (easily around 3000 atoms), current computational power still does not allow for a full, exact-exchange-including DFT treatment. Thus, we here report on the development and benchmarking of an electrostatic embedding QM/MM framework that allows for simulations of systems of these sizes and geometries. The highly arched curvature of spherical NP surface is known to partially dissociate the first water monolayer around it [26,27]. In addition, the curved surface has long-range physical effects on surrounding water, as recently reported [28]. Thus, when preparing simulations of NPs, it is important to assess how much water should be included in the simulation to achieve the desired realism. To describe the aforementioned dissociation of the first monolayer, it is necessary to include it in the QM subsystem in the multiscale divisioning, when working with simple, rigid, point charge-based force fields. This means that there will almost exclusively be water close to the QM/MM intersection, which again motivates a thorough analysis/optimization of the water–water QM/MM coupling accuracy.

The paper is structured as follows: In Section 2, we present the results from the water–water QM/MM LJ re-parameterization and implementation (Section 2.1), before turning to solvation of a $TiO_2$ NP (Section 2.2). In Section 3.1, we analyse the consequences of choosing various LJ fitting strategies, assess the transferability of QM/MM LJ parameters between different electronic structure codes, and the extent of the influence of water–water LJ QM/MM interactions. Section 3.2 discusses the consequences of including more water in the NP simulations. Finally, the computational details of the various sections are presented in Section 4.2, before Section 5 wraps up the conclusions from our work. The appendices contain further benchmarks on our QM/MM implementation.

The updated QM/MM implementation and CRYSTAL17/AMBER16 interface are available in the official version of the Atomic Simulation Environment (ASE) [29,30] (see https://wiki.fysik.dtu.dk/ase/).

## 2. Results

*2.1. Optimizing the Lennard–Jones Parameters for QM/MM Water–Water Interactions*

By comparing QM/MM interaction energies with their pure QM counterparts on a system that only contains one molecular species, we can assess the accuracy of our novel electrostatic embedding framework [11,31]. Since the method is focused on describing molecules and larger systems such as nanoparticles in aqueous solution, we start by focusing on neat water. Some of the main criteria of success for achieving an accurate QM/MM method are that: (1) the QM/MM coupled interaction energies do not over- or under-bind when compared to neither the chosen QM nor the MM models: and (2) the various possible combinations of QM/MM geometries and regions are as similar as possible, so that there will be no orientation-induced artifacts in the total energies and forces [2]. The QM/MM coupling model used in electrostatic embedding (see Section 4.1) leaves open two main routes for the tuning of the accuracy of the coupling: (1) through how the additional electrostatic term in the external potential in the DFT code is constructed [11,32]; and (2) via a re-parameterization of the Lennard–Jones potential that describes non-electrostatic interactions between the atoms in the QM and MM subsystem (Equation (4) in Section 4.1). Here, we focus on the second option. We have employed two strategies for optimizing QM/MM LJ parameters: The *Dimer Fit* (DF) is based on optimizing the LJ parameters to minimize the difference between the multiscale interaction energy curves of the water dimer and the average of the pure QM and pure MM interaction energy curves, a methodology akin to previous work on other systems [33]. The other strategy attempts to take into account that the structure of liquid water cannot be completely understood only analyzing the water dimer, and thus, the *Cluster Fit* (CF) is a global fit on interaction energies on water clusters of size 3–10 molecules per cluster. The details of the fits can be found in Section 4.2.2, which the very interested reader might benefit from reading first, before continuing to the results section. Apart from minimizing the errors in structure and energy introduced by partitioning the system, the fitting effort also leads to insights into the transferability of the molecular characteristics of the water dimer into the modeling of liquid water.

Hybrid functionals provide a better description of the water liquid structure, compared to their GGA relatives [34–36], although often semi-empirical corrections to GGA functionals are employed [37]. In contrast, for the MM subsystem, there are many known improvements to the TIP3P description of water, either within the same general model of static point charges [38], or even extending the description to account for the polarizability of water [39–41], and/or flexible molecules [41,42]. However, employing more advanced MM water potentials is outside the scope of this work because: (1) TIP3P is still ubiquitous in many QM/MM simulations, especially of very large total systems, where the added computational expense of going beyond pairwise potentials can become prohibitive. (2) When including the first solvation shell (or more) in the QM region and focusing on explicit solvation effects on the solute in particular, and not so much the solute–solvent interactions themselves, the outer layers of water become less important, thus making TIP3P an adequate choice.

2.1.1. Water Dimer

Throughout this section, we use the following terminology: "QM/MM" means that the hydrogen-donating molecule is in the QM region, while the hydrogen-accepting is described with TIP3P, and vice versa for "MM/QM". We address both configurations collectively as "the multiscale configurations", to avoid confusion.

Figure 1 reports the structure of the water dimer considered and relevant quantities assessed for our QM/MM implementation.

The hydrogen bonding energy of the water dimer is calculated with the three different sets of LJ parameters, and compared to the pure B3LYP and pure TIP3P results in Figure 2. Using the TIP3P LJ parameters with our choice of functional and basis set within CRYSTAL17 produces overbound multiscale dimer binding curves for both configurations, but most prominently for the QM/MM configuration, where the electronic density of the hydrogen-donating molecule is modeled explicitly.

At very short distances, dominated by Pauli repulsion, the differences in binding energies between the multiscale and the B3LYP curves are large, indicating that a re-evaluation of the LJ parameters are warranted. For the multiscale binding curves using the DF LJ parameters (see Section 4.2.2 for details on the fitting strategies), the QM/MM curve becomes almost identical to the TIP3P result, even though the fit target is the average of the QM and the MM binding curves. The resulting MM/QM curve is also brought closer to its QM counterpart, and all overbinding has been removed from the model. The CF strategy seemingly produces binding curves that are still overbound around their minima, but it interestingly decreases the difference between the O-O distances with the two multiscale minimum values, effectively making the multiscale model more consistent with respect to geometric differences. See Appendix A for an analysis of the various relaxed multiscale geometries for all possible configurations of the water dimer.

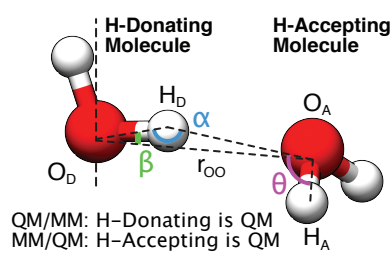

**Figure 1.** Illustration of the geometric quantities used in the water benchmarks. For the dimer tests and fits, we use the shorthand nomenclatures "QM/MM" and "MM/QM" to discern between the two possible subsystem configurations. Three angles provide insight into the hydrogen bonding network structure of liquid water, and are sampled in the molecular dynamics simulations: The donor angles $\alpha = \angle(O_D, H_D, O_A)$ and $\beta = \angle(H_D, O_D, O_A)$, and the acceptor angle $\theta = \angle(H_D, O_A, H_A)$.

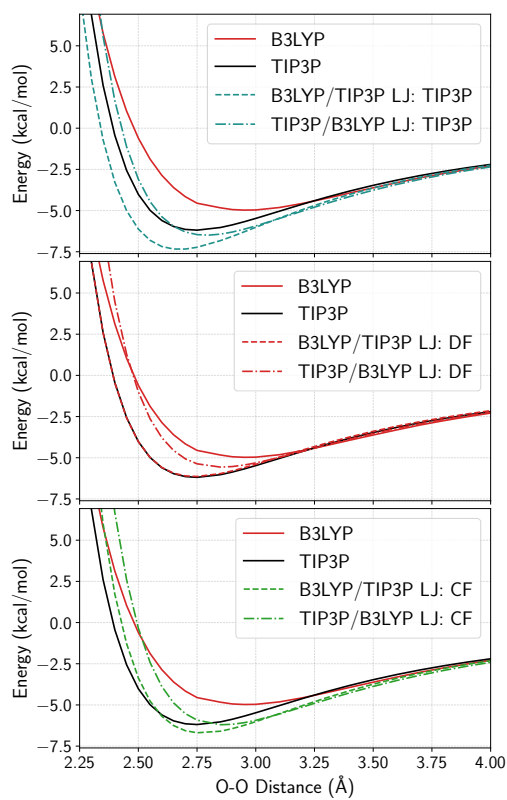

**Figure 2.** Water dimer binding curves (counterpoise corrected), calculated using LJ parameters from: TIP3P (**top**); the dimer fit (**middle**); and the cluster fit (**bottom**). The starting geometry is obtained from the S22 database, where the water dimer was optimized using CCSD(T) [43], and all other intramolecular geometries are kept constant while scanning the O-O distance.

### 2.1.2. Water Clusters

To assess whether the results from the dimer-fitted LJ parameters transfer to larger systems of water, we analyze the difference in interaction QM/MM energies and their pure B3LYP counterparts:

$$\Delta\Delta E = \frac{\Delta E_{\text{int}}^{\text{QMMM}} - \Delta E_{\text{int}}^{\text{B3LYP}}}{n} \quad (1)$$

evaluated over a large dataset of water clusters ranging from $n = 3$ to 10 molecules in total. Since the hydrogen bonding network of liquid water is only fully realized in all three spatial dimensions when tackling clusters containing six or more molecules, including such systems into our fitting methodology seems justifiable in the quest for more transferable QM/MM LJ parameters for water–water interactions. Figure 3 shows box plots representing the distributions in interaction energy differences with respect to the pure B3LYP results. The blue patches represent the maximum difference in interaction energies from the pure B3LYP and pure TIP3P results. Using the TIP3P LJ parameters results in parts of the QM/MM $\Delta\Delta E$-values falling below the blue patches, meaning that there is a significant overbinding with respect to the single-description results. We also observe a trend of the binding systematically being tightest for the multiscale configurations that have the most QM/MM interactions, resulting in "u-shaped" curves. When using the DF LJ parameters, this u-shape has disappeared, but, especially for the 8- and 9-mer clusters, the QM/MM interaction energies give significantly underbound clusters, as the $\Delta\Delta E$-values end up over the upper limit of the blue patches. The CF parameters retain the u-shape, indicative of the same systematic increase in binding with number of QM/MM LJ terms, but all QM/MM distributions are now within the total difference spanned by the limits of the two pure descriptions. In all cases, spread of energy differences is largest when the multiscale description is furthest removed from the reference (i.e., 1 QM water vs. pure QM water), as one would expect for a well-behaved interface. The fact that the TIP3P and CF results are more similar is most likely because the TIP3P parameters are already optimized to reproduce bulk phase properties [2], which is discussed further in Section 3.1.

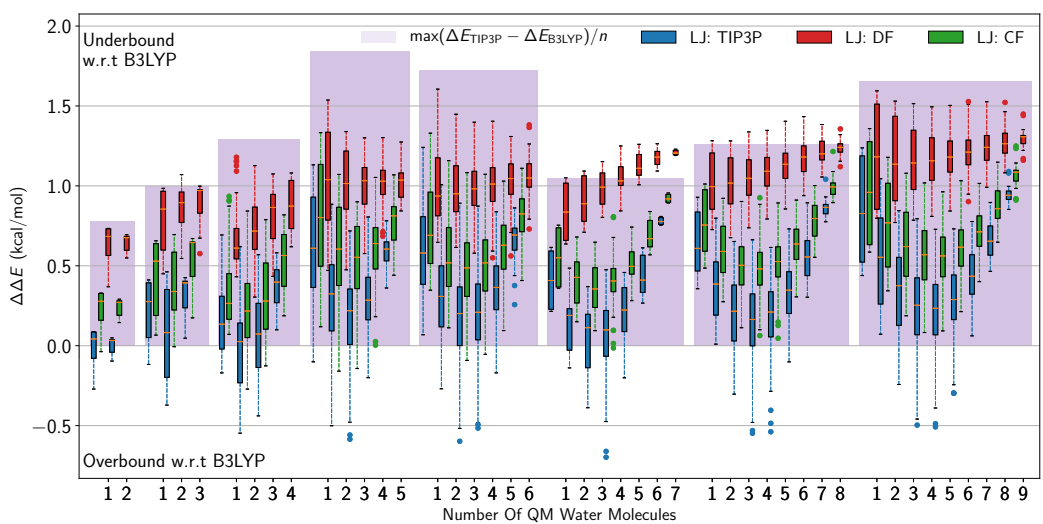

**Figure 3.** Boxplot of the difference in interaction energies with respect to the pure B3LYP results. The horizontal axis is firstly divided divided into eight subsets based on the total number of molecules in each size of water cluster, and then further divided into subsets defined by how many of the total amount of molecules are in the QM subsystem. Each box represents the distribution of interaction energy differences from all possible QM/MM combinations with that number of QM molecules. The boxes spans 50% (the IQR) of the distribution of energy differences, while the whiskers represent ±1.5 IQR. The three datasets are produced using the three different sets of LJ parameters: The original TIP3P parameters (blue), the dimer fit (DF) parameters (red), and the cluster fit (CF) parameters (green). The blue patches behind the box plots show the maximum difference between the two pure descriptions.

## 2.2. Nanoparticles

As mentioned in Section 1, it is important to explore the effects of including increasing amounts of water around NPs, since the nanoparticle/water interfacial effects are quite long range. Here, we have studied three systems: (A) the bare NP $(TiO_2)_{223} \cdot 10H_2O$ (Figure 4a); (B) the NP with a first water monolayer (ML) adsorbed on the surface containing 134 molecules, ~20% of which are dissociated, as discussed and detailed in Section 4.2.5 and in [28] (Figure 4b); and (C), comprised of system B, with a molecular mechanic (MM) region composed by 824 surrounding waters added around it (Figure 4c). The figure shows the structures after geometry optimizations. Systems A and B were originally prepared for a previous study by some of us [28]. The geometry optimization for system C was carried out twice, once with the water in the MM region described with the CF LJ parameters, and once with the TIP3P LJ parameters, to assess any possible effects on the NP.

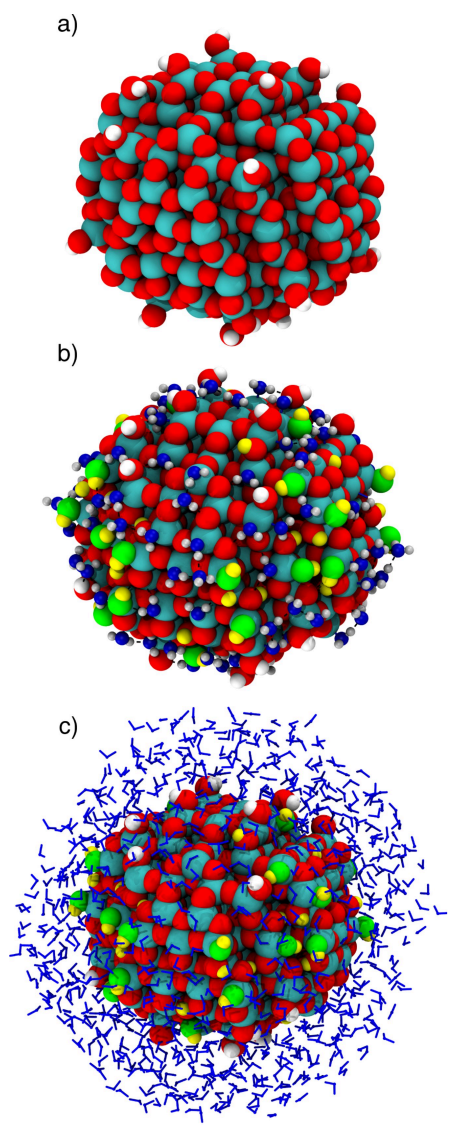

**Figure 4.** Optimized structure with B3LYP: (**a**) for the bare $TiO_2$ nanoparticle; (**b**) for the nanoparticle with an adsorbed water monolayer composed by 134 molecules (~20% of which are dissociated); and (**c**) with QM/MM for the nanoparticle with a water multilayer of 958 molecules, using CF LJ parameters. Titanium, oxygen and hydrogen atoms of the NP are in cyan, red and white, respectively. Oxygen atoms of molecular water are in blue, while oxygen and hydrogen atoms from the dissociated water molecules are in green and yellow, respectively.

First, the effect of water around the NP has been investigated with respect to the induced structural changes in the optimized geometries. In Figure 5, the simulated EXAFS spectra of the bare NP, of the NP with an adsorbed water monolayer and of the one with a water multilayer are compared with that of bulk anatase. The distribution is quite broad for the bare NP, while it is reduced upon water ML adsorption. The change is most drastic going from no water at all to *some* water, but, when the multilayer is added, there is only a further tendency of the distances, in particular of Ti-O$_{ax}$, to be centred at the bulk value, as the grow-in of a new peak at this distance indicates. This is true for calculations performed with both the CF and TIP3P LJ parameters, which produce almost identical spectra.

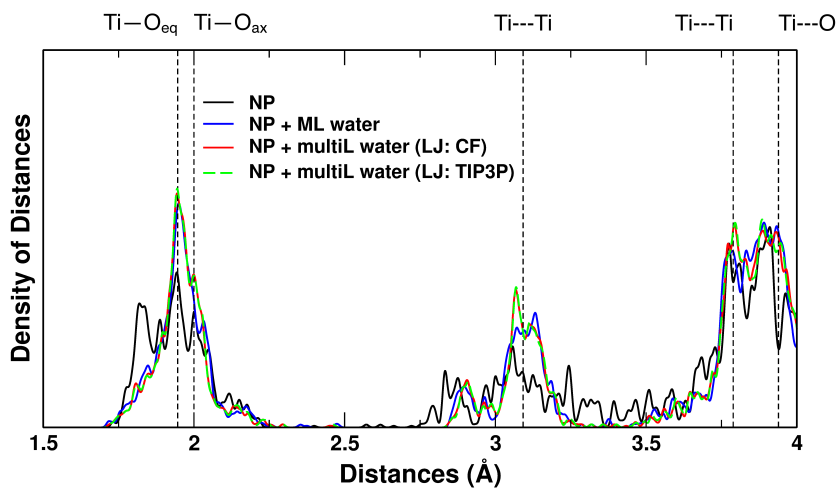

**Figure 5.** Distances-distribution (simulated EXAFS) computed with B3LYP for the bare NP (black), the NP with the adsorbed water monolayer (ML, blue), computed with QM/MM for the NP with the water multilayer using the CF LJ (multiL, red) and using the TIP3P LJ (multiL, dashed green). Dashed black lines correspond to the anatase bulk Ti—O, Ti---Ti and T---O distances.

The effect on the NP electronic properties of the QM/MM relaxation has also been investigated. Figure 6 compares the total (DOS) and projected density of states (PDOS) on oxygen species for the bare NP, the NP with a water monolayer and with a water multilayer around it. The electronic structure of the NP in vacuum has been evaluated both with PBE and B3LYP functionals (Figure 6a,b, respectively). First, when switching from a GGA functional such as PBE to a functional including exact exchange, we observe an expected widening of the bandgap from 2.44 to 4.14 eV [15]. When a ML of water is adsorbed on the NP surface (Figure 6c), we observe a 0.1 eV change in the band gap value, and a decrease of the work function by +0.97 eV. The effect is quite reduced, with a shift of +0.81 eV and +0.83 eV, when the system is immersed in a multilayer of water (Figure 6d), using CF or TIP3P LJ parameters, respectively.

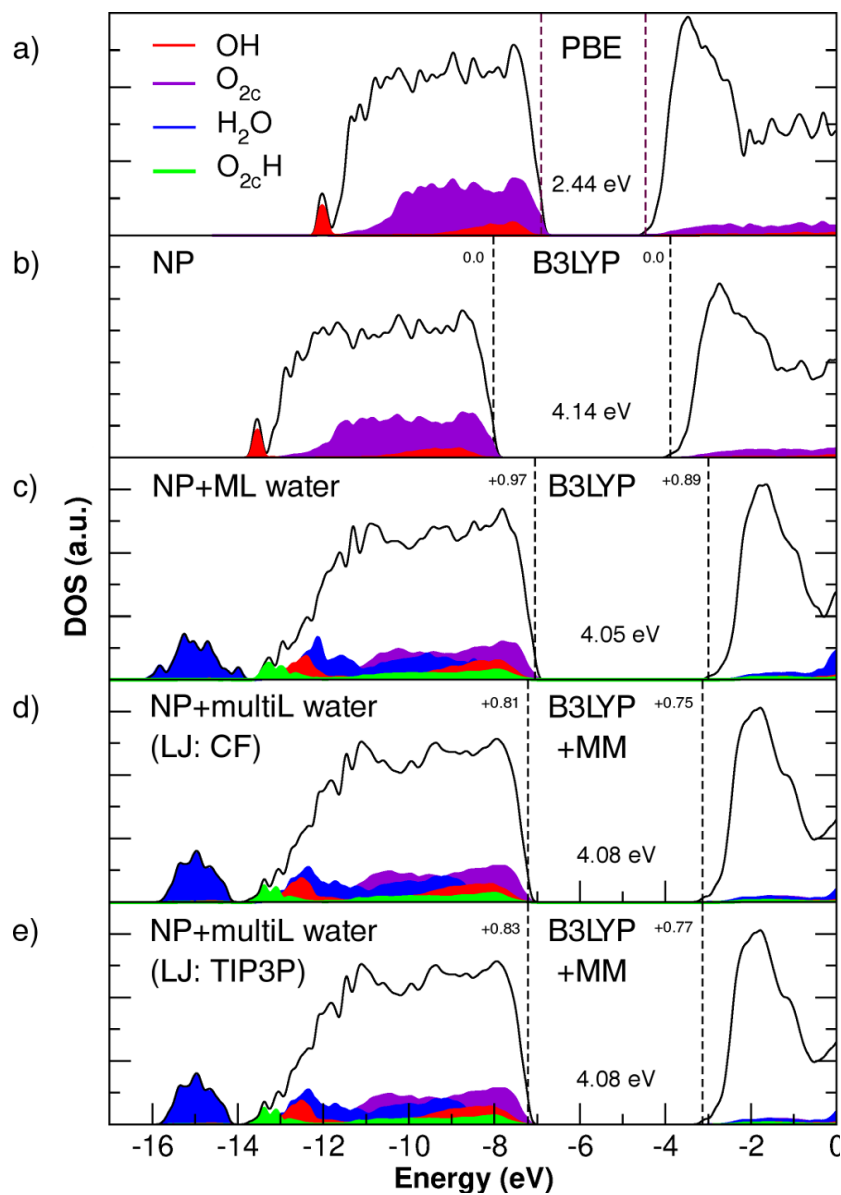

**Figure 6.** Total (DOS) and projected density (PDOS) of states on different oxygen species of: (**a**) the NP in vacuum calculated with the PBE functional; (**b**) the NP in vacuum calculated with the B3LYP functional; (**c**) the NP with a QM water monolayer; (**d**) with a QM/MM water multilayer and the CF LJ parameters; and (**e**) with a QM/MM water multilayer and the TIP3P LJ parameters. In each panel, the band gap and the energy shift of the valence and conduction band edges with respect to the vacuum NP model in a vacuum are given in eV. The zero energy is set to the vacuum level. Panels (**b**,**c**) are from previous work by some of the authors [28].

## 3. Discussion

### 3.1. Water–Water QM/MM LJ Re-Parameterization and Parameter Transferability Tests

Table 1 compares the values of $\sigma_{OO}$ and $\epsilon_{OO}$, used in the non-electrostatic term of the QM/MM coupling (Equation (4) in Section 4.1) obtained from our two different methods of fitting water–water QM/MM LJ parameters, and shows that optimizing LJ parameters only on the water dimer gives rise to a pairwise potential that is very different from both the TIP3P and CF potentials: while the van der Waals (vdW) radius is increased significantly, the potential well depth is over 10 times more shallow. On the contrary, the CF methodology reaches its optimum values by reducing the vdW radius and deepening the potential. The cluster- (Figure 3) and liquid-water benchmarks (Figures A3 and A4 in Appendix C) present a good case for the necessity of reaching beyond dimer systems when generating QM/MM parameters for solvents. This might not be particularly surprising, since the full three-dimensional hydrogen bonding network of water cannot be realized by molecular systems smaller than the water hexamer. Furthermore, the MM parameters are usually optimized to best describe condensed-phase systems that behave systematically more polarized, which again means that MM gas phase potentials might have overemphasized interactions [2]. The description of polarization in the QM system is more transferable between phases, therefore it should not have as overemphasized gas phase interactions as the MM description. Thus, when fitting the QM/MM interaction (partially) to the MM interaction (and partially to the QM interaction which is *not* correspondingly under-interacting), the over-emphasis carries over into the obtained parameters.

**Table 1.** The two sets of QM/MM LJ parameters generated in this work, compared to the TIP3P LJ parameters. See Section 4.1 for their definition.

| Type | $\sigma_{OO}$ (Å) | $\epsilon_{OO}$ (kcal/mol) |
|---|---|---|
| TIP3P | 3.15061 | 0.1521 |
| Dimer Fit (DF) | 3.89048 | 0.0122 |
| Cluster Fit (CF) | 3.10031 | 0.2629 |

Remembering that the DF method with its order of magnitude lower $\epsilon$-value is just short of removing the attractive part of the LJ potential altogether, it is interesting to briefly revisit Figure 3. The two datasets with normal $\epsilon$-values are seen to systematically bind hardest when the number of QM/MM LJ interactions is largest, giving these u-shaped collections of box plots. Since this feature is absent from the DF dataset, we can attribute this type of QM/MM overbinding (within the limits spanned by the pure QM pure MM difference) to the LJ potential. Since performing a fit on the entire dataset of clusters does not remove this behaviour, it must be inherent in the actual formulation of the interaction, and thus, it is necessary to improve on that, or go beyond electrostatic embedding QM/MM altogether, if one wishes to increase the accuracy even further.

It should be noted that when it comes to QM/MM methodologies that involve fitting of one sort or another, many much more "modern" strategies involving machine learning routines have been formulated [6], but how to apply such strategies to explicitly interacting QM and MM subsystems (e.g., electrostatic embedding QM/MM) seems not yet quite realized. The fitting methodology presented here is very basic in comparison, but has the advantage of only requiring re-evaluations of the pairwise LJ potential, which means that, if one were to replace the MM force field, no new QM simulations would have to be made to redo the fit, and, when the new parameters are obtained, there is no extra cost to the following simulations.

In the same vein as our water–water LJ re-parameterization, we have also tested a previous QM/MM re-parameterization of LJ parameters of organic molecule/water dimers to assess the more general accuracy of our novel CRYSTAL17/AMBER16 coupling, the details of which can be found in Appendix B. The QM/MM LJ parameters for this section were obtained from the work of Freindorf et al. When transferring the QM/MM LJ parameters optimized by Freindorf et al. to CRYSTAL17, the RMSD values increase from the results in the paper by 0.11 kcal/mol, 0.01 Å, and 0.98 degrees, for the energy, hydrogen bond distances and angles, respectively. Taking all possible calculational differences mentioned in Appendix B into account, we conclude that simply transferring the previously produced QM/MM parameters to a new QM/MM implementation can be done without a dramatic loss of accuracy. Lastly, we also note that we find no systematicity in any of the errors based on whether the role of the water molecule in the dimers is to act as a hydrogen-donor or -acceptor, which again indicates a robust QM/MM description, which is not significantly and/or systematically sensitive to the nature of the hydrogen bonding geometry over the QM/MM division.

*3.2. Nanoparticles*

For the simulated EXAFS spectra of the three NP systems (Figure 5), we saw that, while the most drastic changes happen when going from naked to ML hydrated NP, there were only fine differences between the QM monolayer and the QM/MM multilayer simulations, in particular associated to the emergence of a peak at the bulk anatase Ti–$O_{ax}$ distance in the QM/MM multilayer simulation. The effects of the QM/MM LJ re-parameterization on the NP structure were negligible, indicating that the inclusion of the first layer of water in the QM subsystem makes the NP structural properties robust to changes in the water–water QM/MM LJ coupling terms. This conclusion is thus also an input to the ongoing debate about the impact of the somewhat ad hoc treatment of non-electrostatic interactions over the QM/MM border. This study seems to show that, as long as this border is far enough removed from the main subject of the study, inaccuracies within the QM/MM non-electrostatic potential are of lesser concern.

The monolayer/multilayer differences could indicate that, when fully immersed in water, the nanoparticle undergoes a slow process of recrystallization, but this would have to be confirmed with molecular dynamics and thermal sampling. An experimental EXAFS of Rajh et al. [44] reported the partial restoration of the octahedral Ti coordination when enediol ligands adsorb to the surface of spherical $TiO_2$ nanoparticles.

With regards to the electronic properties of the NP in different surrounding conditions, it appears evident that inclusion of exact exchange is crucial for providing the correct electronic structure. The exact exchange term partially corrects for the electronic self-interaction error that is present in standard GGA functionals, such as PBE, and causes a large underestimation of the band gap, as is evident from Figure 6. We wish to comment on the fact that the DFTB+/AMBER approach used in previous work [28] is clearly faster than the current approach that does not employ the Tight-Binding approximation. The previous approach allows introducing a much larger number of water molecules to describe the bulk of water around the nanoparticle. However, it cannot correctly capture the electronic properties of the system being based on standard density functional theory methods.

When adding more water, up to a 1 nm thick layer, we do not observe a relevant effect on the band gap, but we observe an effect on the NP's work function with a shift of about 0.2 eV with both CF and TIP3P LJ parameters, due to the dipole orientation of the water molecules in the multilayer. In particular, one should consider that the surface dipole moment for a naked NP is negative and pointing outwards. In the first adsorbed water monolayer (Layer I in Figure 7), all molecules are bound through the water oxygen with the OH bonds directed towards the vacuum. This creates an opposite (positive) dipole moment (i.e., pointing towards the center of the NP), which destabilizes the band states by +0.97 eV (compare the position of the top of the valence band in Figure 6c with that in Figure 6b). The addition of further layers on top of the first adsorbed water monolayer (Layers II–V in Figure 7) mitigates this effect because the average radial dipole moment, created by the multilayer, is negative, causing a shift back of 0.2 eV (compare the position of the top of the valence band of Figure 6d with that in Figure 6c). It must be re-iterated that these observations pertain to a specific configuration of water molecules in the multilayer, as obtained through the relaxation from a starting random distribution (see Section 4.2.5). To assess whether these changes are carried over to a thermal average, one should run molecular dynamics trajectory/ies with extended sampling of configurations, which is however beyond the scope of this work and could be object of future studies.

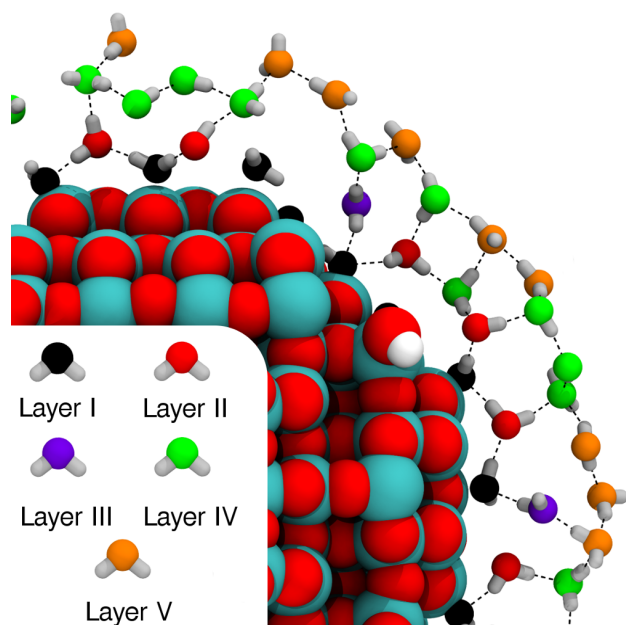

**Figure 7.** Top view of the cross-section of the water multilayer on the NP model after QM/MM geometry optimization. Oxygen atoms of the adsorbed water molecules and OH groups are color-coded according to their distance from the closest titanium atom. Titanium, oxygen, and hydrogen atoms of the nanoparticle are shown as cyan, red, and white spheres, respectively. Hydrogen bonds are represented by dashed black lines.

## 4. Materials and Methods

### 4.1. The Basics of Electrostatic Embedding QM/MM

In electrostatic embedding QM/MM methodologies, the total energy of the entire system is a sum of the two subsystems, plus an interaction term:

$$E_{\text{TOT}} = E_{\text{QM}} + E_{\text{QM/MM}} + E_{\text{MM}} \tag{2}$$

The MM region is here described using a point charge-based force field, with $q_i$ denoting the charges and $\mathbf{R_i}$ their spatial coordinates and $n(\mathbf{r})$ the electronic density of the QM subsystem. The interaction energy term $E_{\text{QM/MM}}$ (in atomic units) is:

$$E_{\text{QM/MM}} = \sum_{i=1}^{N_{\text{MM}}} q_i \int \frac{n(\mathbf{r})}{|\mathbf{r}-\mathbf{R_i}|} d\mathbf{r} + \sum_{i=1}^{N_{\text{MM}}} \sum_{\alpha=1}^{N_{\text{QM}}} \frac{q_i \mathcal{Z}_\alpha}{|\mathbf{R}_\alpha - \mathbf{R_i}|} + E_{\text{NES}} \tag{3}$$

with $\mathcal{Z}_\alpha$ being the atomic number of atom $\alpha$, running over all atoms in the QM subsystem. $i$ runs over all $N_{\text{MM}}$ charges of the MM subsystem. In CRYSTAL17, the point charges are then formally considered "atoms" with zero mass, nuclear charge as given in the input, and the default assigned basis set is a single $s$ gaussian with an exponent of 1,000,000. Thus, the point charges are included in the external potential, and the electronic density is converged under their influence. Likewise, CRYSTAL17 calculates the electrostatic forces from the density of the QM subsystem on the point charge centres.

The non-electrostatic electronic interactions, $E_{\text{NES}}$, are in this work included through a Lennard–Jones pair potential between the atomic centers of each subsystem:

$$E_{\text{NES}} = \sum_{i}^{N_{\text{MM}}} \sum_{\alpha}^{N_{\text{QM}}} 4\epsilon_{i\alpha} \left[ \left( \frac{\sigma_{i\alpha}}{|\mathbf{R}_\alpha - \mathbf{R}_i|} \right)^{12} - \left( \frac{\sigma_{i\alpha}}{|\mathbf{R}_\alpha - \mathbf{R}_i|} \right)^{6} \right] \quad (4)$$

with $\epsilon_{i\alpha}$ defining the well-depth, and $\sigma_{i\alpha}$ the equillibrium distance of the potential. Thus, no non-electrostatic interactions are directly taken into consideration in the evaluation of the electronic density of the QM subsystem. The re-parameterizations via the DF and the CF strategies are carried out for $\epsilon_{i\alpha}$ and $\sigma_{i\alpha}$ of Equation (4), for the atom types (elements) Oxygen and Hydrogen, combined via the Lorenz–Berthelot combining rules.

*4.2. Computational Details*

For the QM subsystems in all of the QM/MM simulations carried out for this work, the choice of functional has been (the CRYSTAL17 implementation of) B3LYP, comprised of Becke's exchange functional [45], 20% HF exchange, and the Lee-Yang-Parr correlation functional [46]. For the benchmarks, we opted for the CRYSTAL17 basis set 6-31G(d,p) [47], which resembles the basis set used in the work where we obtained the non-water QM/MM LJ parameters from [33]. The Anderson Convergence Accelerator was used for all CRYSTAL simulations [13], and the CRYSTAL standard convergence criteria for single point calculations were used. For water simulations including more than a single QM water molecule, the counterpoise correction scheme was employed to avoid basis set superposition errors: when calculating monomer energies of single molecules in clusters, ghost atoms were added at the positions of the other molecules, so that the same quality of basis set was kept constant, when evaluating interaction energies. As in previous work [11], the ASE-native TIP3P implementation [30,48] was used for the pure water benchmarks, thus avoiding the process of generating AMBER input files for these 6000+ calculations.

4.2.1. Water Dimers and Clusters

The interaction energy, or binding energy, is defined as $\Delta E_{\text{int}} = E_{\text{cluster}} - \sum_{m}^{n} E_m$, where $E_m$ is the energy of monomer $m$ in the cluster, and $n$ is the total number of molecules in the cluster. Thus, e.g., for a QM/MM water dimer using TIP3P as the MM model, only the QM monomer has a monomer energy.

To transferably and generally minimize over- or underbinding of QM/MM systems in relation to the total difference between pure B3LYP and TIP3P models, we go beyond the water dimer, and extend the water dataset to encompass binding energy calculations of all $N$ possible QM/MM combinations of $n$-molecule water clusters:

$$N = \sum_{n=3}^{10} \sum_{k=1}^{n-1} \frac{n!}{k!(n-k)!} c_n \quad (5)$$

where $k$ is the number of QM molecules, and $c_n$ is the number of clusters with $n$ total molecules in the dataset. The water cluster geometries were obtained from the work by the Bates and Tschumper [49], and the set of $(H_2O)_n$, $n = 3$–10 clusters provided by Temelso and coworkers [50], for a total of 6214 QM/MM single point energy calculations. First, a dataset of interaction energy differences $\Delta\Delta E = \Delta E_{\text{int}}^{\text{QMMM}} - \Delta E_{\text{int}}^{\text{B3LYP}}$ was produced using the TIP3P LJ parameters. The resulting dataset will then consist of a distribution of interaction energies per $k$, for each $n$. This means quite a few distributions will have to be held up against each other, which is most easily visualized through a box-plot, where the distribution is simplified by showing the interquartile range (IQR), or the middle 50% of the distribution as a box, as done in Figure 3.

Since only electrostatic interactions are included in the external potential in CRYSTAL17, one can simply subtract the non-electrostatic contributions to the interaction energies, and re-calculate the LJ term with any given LJ parameters, to achieve new total interaction energies, without having to redo the more expensive DFT calculations. This method was employed to obtain fitted LJ parameters that minimized the total the over- and underbinding of the entire water cluster dataset, as described in the following section.

4.2.2. Fitting Strategies

The dimer fit minimizes the difference between the QM/MM dimer binding curve and the average of the pure QM and the pure MM also in the 2.1 < $r$ < 3.0 Å interval, by optimizing the LJ parameters that are used in the QM/MM non-electrostatic coupling term (Equation (4)). The MM-MM LJ interactions are not modified, i.e., these interactions are still evaluated using the TIP3P parameters. The fit was done using the curve_fit tool from the Scipy.optimize submodule for python. The interval is chosen to avoid giving weight to the very repulsive region at $r$ < 2.1 Å, which would skew the overall interaction.

The cluster fit minimizes a modified chi-square difference, $\chi^2$, summed over all $i$ cluster configurations, and normalized with the number of molecules in each cluster $N_i$:

$$\chi^2 = \sum_i \frac{\left(\Delta E_{\text{int},i}^{\text{QMMM}} - \Delta E_{\text{int},i}^{\text{avg}}\right)^2}{N_i \Delta E_{\text{int},i}^{\text{avg}}} \quad (6)$$

where $\Delta E_{\text{int},i}^{\text{avg}}$ is the average of the pure QM and pure MM interaction energy of cluster configuration $i$. The fit was carried out by using the basin hopping algorithm implemented in the Scipy python module, starting from TIP3P LJ parameters, but also allowing for non-zero LJ parameters on the hydrogen atoms, for a total of 4 degrees of freedom in the fit. The initial step size was $1 \times 10^{-4}$ eV or Å. Keeping in mind the physical meaning of the $\sigma$- and $\epsilon$-parameters in the LJ potential, the following constraints were employed: 1 Å < $\sigma_{\text{OO}}$ < 6 Å, and 0 Å ≤ $\sigma_{\text{HH}}$ < 3 Å. After this initial exploration of the parameter space, a bootstrapping methodology was employed, spanning an initial grid of all four free parameters within the limits of the constraints. The minimizer used here was of the sequential least squares programming type, also from the scipy.minimize module. This provided the smallest $\chi^2$, with the parameters $\sigma_{\text{OO}}^{\text{CF}}$ = 3.10031 Å, $\epsilon_{\text{OO}}^{\text{CF}}$ = 0.011402 eV, and curiously no non-zero hydrogen LJ parameters.

4.2.3. Liquid Water Simulations

In preparation for the QM/MM liquid water simulations, a cubic box with 27.96 Å sides containing 729 water molecules was equilibrated at 300 K with the TIP3P force field, using a Langevin thermostat with a friction coefficient of 0.02 a.u. After the equilibration was completed, the simulation was continued to produce initial configurations for spawning QM/MM MD simulations at 1 ps intervals from the equilibrated MM MD trajectory. Three sets of QM/MM MD simulations were prepared: (1) using the TIP3P LJ parameters; (2) using the DF LJ parameters; and (3) using the CF parameters. The simulation setups were otherwise identical: The QM subsystem was defined to include a single water molecule, and re-equilibration was carried out for 4 ps for each of the spawned QM/MM MD trajectories before sampling of, e.g., RDFs, was carried out. The RDFs were calculated using the VMD program [51]. The total sampled time for the CF LJ parameter-runs were 182 ps, 123 ps for the DF LJ parameter-runs, and 56 ps for the TIP3P runs. The results can be found in Appendix C.

4.2.4. Organic Molecule/Water Dimer Benchmarks

To benchmark the accuracy of our QM/MM methodology against pure DFT results on organic molecules, we used a subset of the database used by Freindorf et al. [33]. All dimers consisting of a single organic molecule and a water molecule from the original database were recreated, as well as two new systems, namely glycine and aspartic acid, were included in our dataset. Their geometries were then optimized with B3LYP/6-31G(d,p) only. Then, following Freindorf et al., QM/MM geometry optimizations were performed, with the MM subsystem consisting of the water molecule, using the original TIP3P LJ parameters, since the new parameters are only optimized for QM/MM water–water interactions. In these calculations, the MM atoms were fixed at their positions obtained from the previous calculations of the dimers, performed with the single-description QM level of theory. The geometry relaxations were carried out until the maximum force magnitude within the systems was below 0.01 eV/Å. The results can be found in Appendix B.

4.2.5. Nanoparticle Simulations

The anatase $TiO_2$ spherical nanoparticle (NP) model used in this work was obtained from a simulated annealing process described in a previous work of some of us [24]. The model has a stoichiometry of $(TiO_2)_{223} \cdot 10H_2O$ and it is characterized by a diameter of about 2.2 nm (see Figure 4a). A water monolayer around the NP was modeled by adding to the undercoordinated Ti atoms of the surface water molecules. The procedure is described in detail in ref. [28]. The most stable water monolayer around the NP is constituted by 134 water molecules and has an extent of dissociation $\alpha = 0.21$ (see Figure 4b). $\alpha$ is defined as:

$$\alpha = \frac{n_{\text{OH,H}}}{n_{\text{tot}}}, \qquad n_{\text{tot}} = n_{\text{H}_2\text{O}} + n_{\text{OH,H}} \tag{7}$$

where $n_{\text{H}_2\text{O}}$ is the number of molecular and $n_{\text{OH,H}}$ of dissociated water molecules in the water monolayer. Finally, a spherical water shell of about 1 nm thickness was added around the NP with the water monolayer using the PACKMOL code [52]. The water density inside the shell is approximately of 1 g/cm$^3$ corresponding to 824 water molecules (see Figure 4c). For all the DFT calculations in this section, the all–electron basis sets used are O 8–411(d1), Ti 86–411(d41), and H 511(p1), as defined in [23]. In the case of the bare NP and the NP with the water monolayer, forces were relaxed to less than 0.02 eV/Å using the CRYSTAL17 code. For the bare NP, we also used the generalized gradient approximated (GGA) functional PBE [53], for comparison to B3LYP results. In the case of the NP surrounded by the multilayer, forces were relaxed to 0.05 eV/Å to decrease the overall computational cost, using CRYSTAL17/AMBER16 QM/MM approach. For the QM part, we used the B3LYP functional, while to describe the QM/MM non-electrostatic interaction we made two simulations, one using CF and one using TIP3P LJ parameters.

To simulate the direct-space extended X-ray adsorption fine structure (EXAFS) spectra, Gaussian convolution of peaks ($\sigma = 0.005$ Å) was centred at the distance lengths between each Ti atom and other atoms (O or Ti) from its first, second, and third coordination shells. Reference anatase bulk Ti-O representative distances have been calculated with DFT(B3LYP) to be, Ti—$O_{eq}$ = 1.946 Å and Ti—$O_{ax}$ = 2.000 Å for the first coordination sphere, Ti- - -Ti = 3.092 Å for the second coordination sphere and Ti- - -Ti = 3.789 Å and Ti- - -O = 3.939Å for the third and fourth coordination sphere, respectively.

Through another Gaussian convolution, this time of the eigenvalue coefficients ($\sigma = 0.005$), centred at the Kohn–Sham energy eigenvalue of each orbital, we simulated the total density of states (DOS) of the nanoparticles. For the projected density of states (PDOS), we used the coefficients in the linear combination of atomic orbitals (LCAO) of each molecular orbital, since summing the squares of the coefficients of all the atomic orbitals centred on a certain atom type results (after normalization) in the relative contribution of each atom type to a specific eigenstate.

## 5. Conclusions

With this work, we have made a QM/MM electrostatic embedding implementation that combines (MPP-)CRYSTAL17 and AMBER16, and tested it thoroughly. We have produced new water–water QM/MM LJ parameters for B3LYP/TIP3P, which improve the water–water coupling over the QM/MM border, and could help reduce problems in, e.g., future adaptive QM/MM methodologies, where the contents of the QM subsystem is dynamically updated. Through this part of the study, we have also made it clear that for solvents such as water, where the water–water interactions over the border are important, it is not always sufficient to optimize new LJ parameters from dimer interactions. This is of extra importance when the fit includes matching an average of QM and MM potentials, where the latter is already optimized to reproduce bulk phase properties, and thus have overemphasised gas-phase interactions.

We have also analysed the transferability of LJ parameters between different DFT codes and QM/MM implementations, and found that the QM/MM vs. pure-QM errors did not increase significantly, even though the parameters were produced with a different DFT code.

Finally, we have demonstrated that the implementation can handle geometry optimizations of $TiO_2$ nanoparticles immersed in water. In the same vein of assessing how large a role the QM/MM LJ parameters play in the total picture, we have shown that changing them has a negligible effect on the NP. This is perhaps not surprising, since the differences do not directly involve the NP itself: the MM-to-QM-water forces would have to affect the QM water shell structure rather drastically for this to affect the NP.

In addition, as the studies on neat QM/MM water have shown, if attempts to reduce computational costs further were to be made by describing *all* water layers with force fields, it is likely that re-parameterization of Ti and NP oxygen–water LJ parameters would be necessary, apart from also having to use a force field for water that can model dissociation.

Our analysis showed that the major changes to the nanoparticle take place already after adding the first water monolayer, and that adding further water layers can have a small effect on the structural properties, here seen in terms of a finer recrystallization. For the specific water configurations studied here, we observed a more evident effect on the electronic properties of the NP. Future work could be focused on studying if and how these effects could change with thermal sampling of the solvated nanoparticles.

**Author Contributions:** D.S., G.F. and J.J.M. wrote the ASE interface to CRYSTAL17, with assistance from A.O.D. and B.C. A.O.D. carried out the water benchmarks and the water-LJ QM/MM fitting work. L.F. and D.S. carried out the benchmark on organic molecule/water dimers. D.S. and C.D.V. carried out the NP simulations, and wrote the sections of the paper pertaining to those. B.C. wrote the details of the charge embedding in CRYSTAL17. A.O.D wrote the rest of the paper. C.D.V. edited the manuscript and coordinated the work.

**Funding:** This research was funded by the Icelandic Research Fund (grant 174244-051) and VILLUM FONDEN, the European Research Council (ERC) under the European Union's HORIZON2020 research and innovation programme (ERC Grant Agreement No [647020]).

**Acknowledgments:** A.O.D. Would like to thank Jónsson, H. for discussions about fitting strategies. C.D.V. is grateful to Lara Ferrighi, Massimo Olivucci, and Stefano Motta for fruitful discussions. A.O.D. Acknowledges funding from the Icelandic Research Fund (grant 174244-051) and VILLUM FONDEN. The project has received funding from the European Research Council (ERC) under the European Union's HORIZON2020 research and innovation programme (ERC Grant Agreement No [647020]).

**Conflicts of Interest:** The authors declare no conflict of interest.

## Abbreviations

The following abbreviations are used in this manuscript:

ADF     Angular Distribution Function
AIMD   Ab Initio Molecular Dynamics
BOMD   Born–Oppenheimer Molecular Dynamics

| CF | Cluster Fit |
|---|---|
| DF | Dimer Fit |
| DOS | Density of States |
| GGA | Generalized Gradient Approximation |
| HF | Hartree-Fock |
| IQR | Interquartile Range |
| LCAO | Linear Combination of Atomic Orbitals |
| LJ | Lennard–Jones |
| ML | Monolayer |
| multiL | Multilayer |
| NP | Nanoparticle |
| PDOS | Projected Density of States |
| QM/MM | Quantum Mechanical/Molecular Mechanical |
| RDF | Radial Distribution Function |
| RMSD | Root Mean Square Deviation |
| vdW | van der Waals |

**Appendix A. Further Water QM/MM Benchmarks**

**Table A1.** Fully optimized geometries of the water dimer, using the various multiscale- and single-description configurations, and their corresponding interaction energies. The QM atoms were completely unconstrained while the MM atoms where constrained to the rigid TIP3P geometry. The optimizations were carried out until the maximum force component on any unconstrained atom was below 0.01 eV/Å.

| Configuration | $r_{OO}$ (Å) | $\angle(\alpha)$ (deg.) | $\angle$(OOH) (deg.) | $\Delta E_{int}$ (kcal/mol) |
|---|---|---|---|---|
| B3LYP | 2.8803 | 51.7 | 6.3 | −5.08 |
| B3LYP/TIP3P (LJ: TIP3P) | 2.6414 | 15.9 | 0.4 | −8.76 |
| B3LYP/TIP3P (LJ: DF) | 2.7039 | 21.6 | 2.0 | −7.31 |
| B3LYP/TIP3P (LJ: CF) | 2.7414 | 17.8 | 0.4 | −7.84 |
| TIP3P/B3LYP (LJ: TIP3P) | 2.8578 | 61.3 | 0.1 | −6.51 |
| TIP3P/B3LYP (LJ: DF) | 2.8502 | 72.2 | 3.2 | −5.49 |
| TIP3P/B3LYP (LJ: CF) | 2.7859 | 61.0 | 0.1 | −6.22 |
| TIP3P | 2.7461 | 20.3 | 4.4 | −6.54 |
| CCSD(T)/cc-pVQZ [a] | 2.9104 | 60.3 | 4.8 | −5.02 |

[a] Multireference calculations data from the S22 dataset [43].

Table A1 shows results from performing geometry optimizations using the three different LJ parameter sets. Note how the multiscale geometries most closely resemble each of the two single-model geometries depending what model is used to describe the lone pair-donating (hydrogen accepting) molecule in the water dimer. The CF LJ parameters provide the smallest difference in optimal O-O distance between the QM/MM and MM/QM configuration.

**Appendix B. Organic Molecule/Water Dimer Benchmark**

*Appendix B.1. Structures and Nomenclature*

Figure A1 provides a visual overview of the benchmarked dimers, and the hydrogen-bonding geometries that the QM/MM interface needs to describe.

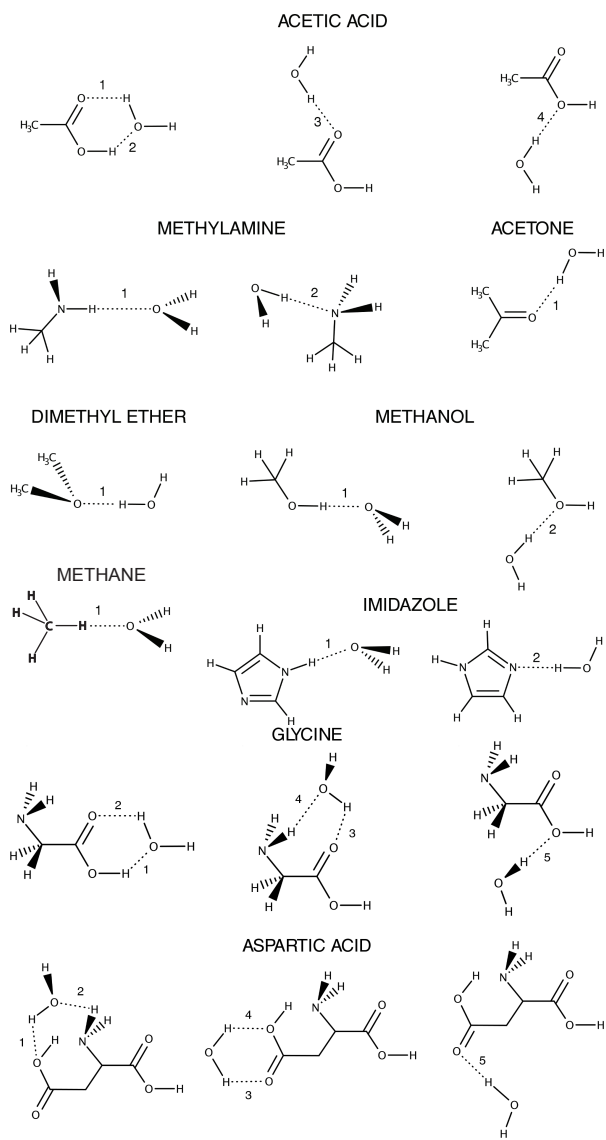

**Figure A1.** Structure and nomenclature of the different H-bond evaluated for different organic molecules/water pairs: acetic acid/water (1,2,3,4), methylamine/water (1,2), acetone/water (1), dimethyl ether/water (1), methanol/water (1,2), methane/water (1), imidazole/water (1,2), glycine/water (1,2,3,4,5), and aspartic acid/water (1,2,3,4,5). It is debatable whether the methane–water interaction can be called a hydrogen bond, but, since it is included in the original dataset, we also include it here.

*Appendix B.2. Benchmark Results*

**Table A2.** Comparison of the hydrogen bond binding energy ($E_{HB}$ in kcal/mol), distance ($r_{HB}$ in Å) and angle ($\phi_{HB}$ in degrees) as calculated with B3LYP and B3LYP/TIP3P. For the structure and nomenclature of the hydrogen bond evaluated for different organic molecules/water and amino acids/water pairs refer to Figure A1 in Appendix B.1.

| Molecule | HB | B3LYP | | | B3LYP/TIP3P | | |
|---|---|---|---|---|---|---|---|
| | | $E_{HB}$ | $r_{HB}$ | $\phi_{HB}$ | $E_{HB}$ | $r_{HB}$ | $\phi_{HB}$ |
| Acetic Acid | 1 | 10.50 | 1.782 | 156.9 | 9.34 | 1.972 | 160.8 |
| | 2 | 10.50 | 1.979 | 136.0 | 9.34 | 2.198 | 135.0 |
| | 3 | 5.94 | 1.930 | 160.0 | 6.03 | 1.987 | 160.2 |
| | 4 | 3.58 | 2.067 | 153.6 | 3.97 | 2.081 | 153.7 |

**Table A2.** *Cont.*

| Molecule | HB | B3LYP | | | B3LYP/TIP3P | | |
|---|---|---|---|---|---|---|---|
| | | $E_{HB}$ | $r_{HB}$ | $\phi_{HB}$ | $E_{HB}$ | $r_{HB}$ | $\phi_{HB}$ |
| Methyl Amine | 1 | 2.68 | 2.188 | 158.7 | 3.20 | 2.175 | 166.3 |
| | 2 | 7.90 | 1.908 | 171.2 | 6.59 | 2.055 | 173.4 |
| Acetone | 1 | 6.31 | 1.907 | 164.6 | 6.47 | 1.948 | 164.9 |
| Dimethyl Ether | 1 | 5.24 | 1.907 | 175.6 | 5.05 | 1.952 | 166.1 |
| Methanol | 1 | 5.58 | 1.936 | 173.9 | 5.93 | 1.946 | 172.4 |
| | 2 | 5.96 | 1.898 | 177.3 | 6.04 | 1.932 | 176.8 |
| Imidazole | 1 | 6.50 | 1.953 | 179.2 | 6.77 | 1.964 | 178.5 |
| | 2 | 7.08 | 1.949 | 179.4 | 6.05 | 2.092 | 177.0 |
| Methane | 1 | 0.38 | 2.566 | 156.7 | 0.74 | 2.660 | 163.6 |
| Glycine | 1 | 10.50 | 1.776 | 156.6 | 9.29 | 1.959 | 160.3 |
| | 2 | 10.50 | 1.995 | 135.0 | 9.29 | 2.204 | 134.4 |
| | 3 | 6.34 | 1.943 | 156.8 | 7.05 | 2.019 | 156.7 |
| | 4 | 6.34 | 2.186 | 146.1 | 7.05 | 2.167 | 156.0 |
| | 5 | 3.26 | 2.045 | 152.2 | 3.80 | 2.101 | 152.9 |
| Aspartic acid | 1 | 6.56 | 2.155 | 135.1 | 7.03 | 2.321 | 134.1 |
| | 2 | 6.56 | 2.563 | 111.2 | 7.03 | 2.820 | 108.1 |
| | 3 | 5.66 | 2.178 | 149.1 | 6.70 | 2.167 | 148.8 |
| | 4 | 5.66 | 2.477 | 126.0 | 6.70 | 2.479 | 126.2 |
| | 5 | 7.29 | 1.893 | 159.9 | 7.26 | 1.968 | 159.8 |

**Table A3.** Root-mean-square deviations (RMSD) for hydrogen bond binding energies ($E_{HB}$ in kcal/mol), distances ($r_{HB}$ in Å), and angles ($\phi_{HB}$ in degrees) between the pure B3LYP reference and the B3LYP/TIP3P calculations.

| RMSD ($E_{HB}$) | RMSD ($r_{HB}$) | RMSD ($\phi_{HB}$) |
|---|---|---|
| 0.77 | 0.121 | 3.7 |

*Appendix B.3. Analysis*

Figure A2 shows the correlation of hydrogen bond energies, distances, and angles between the QM/MM simulations and the pure QM reference. The QM/MM angles and energies are evenly distributed around their QM reference, and no systematic differences can be observed for this dataset. For the hydrogen bond lengths, the correlation still seems linear, but with a tendency for the QM/MM model to systematically slightly overestimate the bond lengths. The total RMSD values for the dataset are 0.77 kcal/mol, 0.121 Å, and 3.7 degrees, for the binding energy, H-bond length, and H-bond angle, respectively.

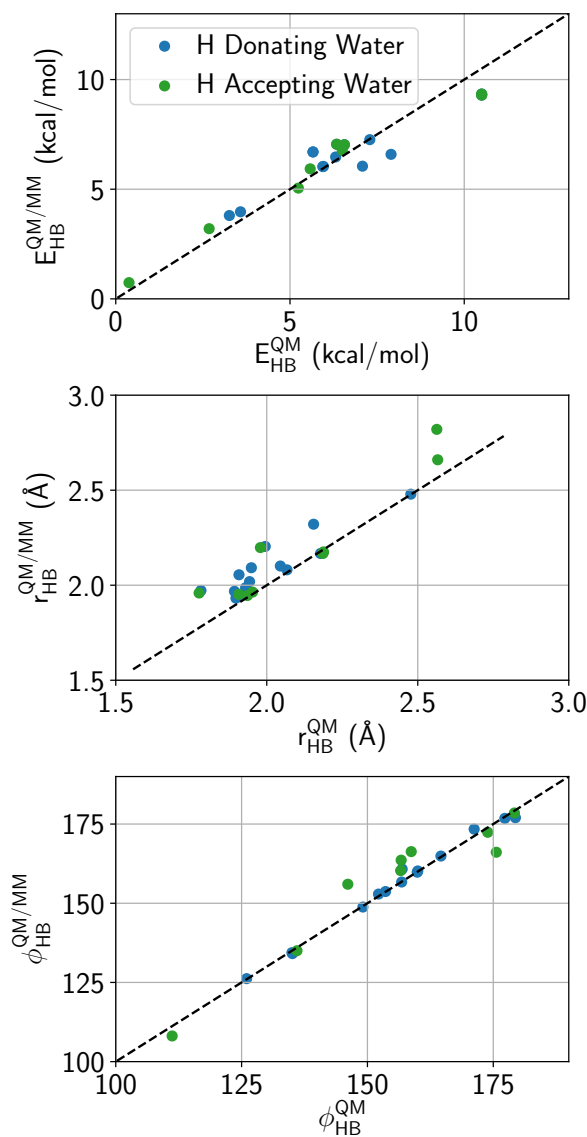

**Figure A2.** QM/MM vs. QM Correlations between hydrogen bond: binding energy (**top**); distance (**middle**); and angle (**bottom**). The dashed black lines represent the pure QM reference values.

The authors report that their hydrogen bond energies are systematically too big, while the hydrogen bond distances are systematically too short [33]. Curiously, when we use their LJ parameters in our CRYSTAL17/AMBER16 QM/MM implementation, we only see, as shown in Figure A2, a systematic deviation of the hydrogen bond distances, and, in our case, the QM/MM bonds are too *long*. There are three main differences between the work of Freindorf et al. and the work carried out here: firstly, the DFT code used by Freindorf et al. is Q-Chem, which is different from CRYSTAL17 in many aspects, and could have an implementation of B3LYP that is not exactly the same as in CRYSTAL17. The authors did not mention if any reduced-scaling algorithms were employed, which we assume must mean that no such methods were used. However, some programs apply, e.g., the resolution of identity approximation by default, which could affect geometry optimizations, but, since no auxiliary basis sets are reported, this is most likely not the case. Secondly, there is not a complete 1:1 correspondence between the basis sets available to the two codes, e.g., CRYSTAL17 uses 5d functions in the 6-31G(d,p) basis set, where many other codes would use 6d. Lastly, the geometry optimizations could have been carried out with another type of optimization, and to different convergence criteria.

**Appendix C. QM/MM Liquid Water**

Figure A3 compares QM/MM water RDFs to the RDF obtained from a pure TIP3P run. Since hybrid functionals such as B3LYP are still relatively costly, we have not made a pure B3LYP simulation for comparison. Instead, we have shown a result from the literature [36], which was restricted to 32 molecules and short trajectories due to cost, and run at 350 K, which impedes direct comparison. It is a well known strategy [54–56] to increase the temperature in AIMD simulations to counteract both the overstructuring of liquid water by many DFT functionals, and to nullify the lack of proton quantum effects in the dynamics. The overbinding in the TIP3P LJ parameter dimer binding curve is observed to carry over to liquid water, and results in an overstructuring: the amplitude of both the first peak, and the first well is the largest of all the datasets. The first peak of the DF and CF datasets are almost of the same height, with a peak magnitude of 2.92 and 2.86 for the CF and DF set, respectively. Both the first well and the second peak of the DF dataset have higher amplitudes than the CF dataset. The CF radius of the first solvation shell seems to lie in between the pure B3LYP and pure TIP3P result, whereas the shell is thinner for the two other QM/MM simulations.

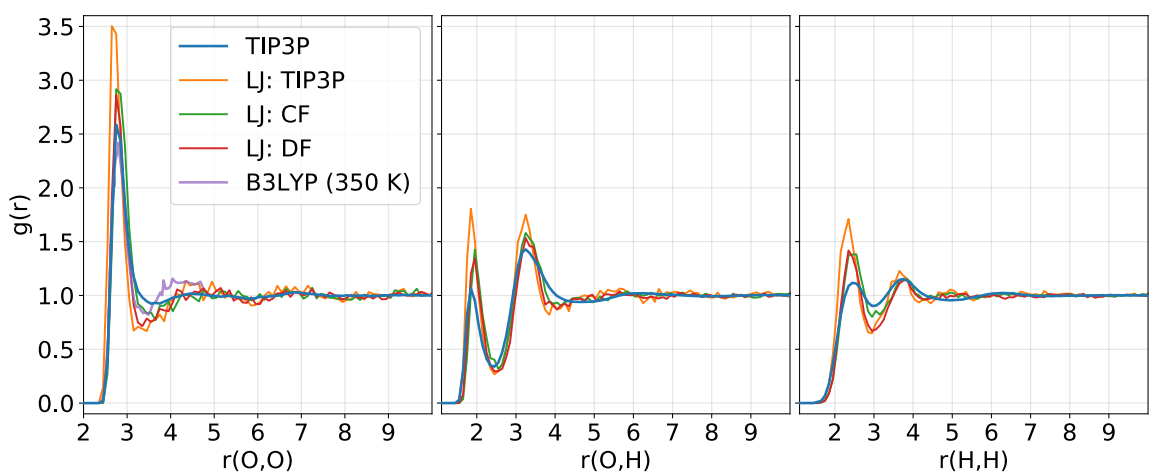

**Figure A3.** Pairwise O-O, O-H, and H-H radial distribution functions of liquid water at 300 K. The QM/MM RDFs are sampled from a single B3LYP water molecule in a box of 728 MM molecules. Three types of QM/MM simulations were made, one for each of the QM/MM LJ parameter sets: TIP3P, DF, and CF. The pure B3LYP data, simulated at 350 K, and for 32 water molecules was obtained from [36]. The higher temperature is known to lower the amplitude of the O-O peaks for both classical, pairwise force fields [54], and for DFT methodologies [55,56].

Angular distributions of the hydrogen bond accepting and donating water molecules are central quantities for characterizing the hydrogen bond network in liquid water. Figure A4 shows the sampled distributions for the three QM/MM water datasets produced in this work. Again, as was the case for the radial distributions, the dimer fitting methodology has reduced the water over-structuring relative to the TIP3P-structure, but not to the same degree as the cluster fit methodology. Especially for the $\alpha$ (and the connected $\beta$) angle, not only the amplitude has been decreased towards the TIP3P result, but also the peak position has moved.

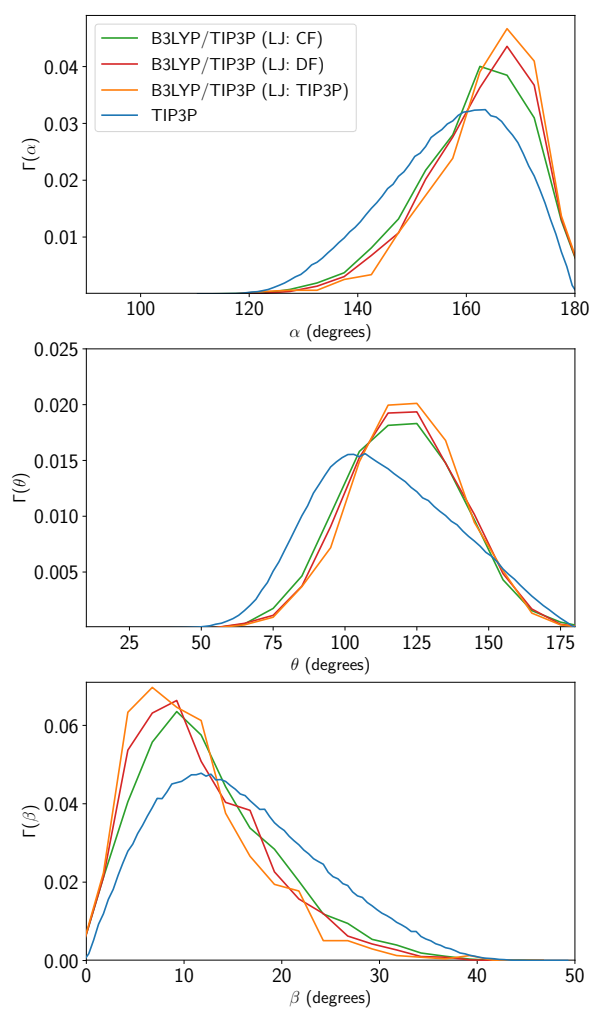

**Figure A4.** QM/MM and pure MM angular distribution functions. See Figure 1 for definitions. The *α* and *β* angles are most affected by the change of QM/MM LJ parameters to the CF set, whereas the DF parameters provide angular orientations that are similar to the structure from the TIP3P parameters.

**References**

1. Warshel, A.; Levitt, M. Theoretical Studies of Enzymatic Reactions: Dielectric, Electrostatic and Steric Stabilization of the Carbonium Ion in the Reaction of Lysozyme. *J. Mol. Biol.* **1976**, *103*, 227–249. [CrossRef]
2. Lin, H.; Truhlar, D.G. QM/MM: What have we learned, where are we, and where do we go from here? *Theor. Chem. Acc.* **2007**, *117*, 185–199. [CrossRef]
3. Senn, H.M.; Thiel, W. QM/MM methods for biomolecular systems. *Angew. Chem. Int. Ed.* **2009**, *48*, 1198–1229. [CrossRef] [PubMed]
4. Bulo, R.E.; Michel, C.; Fleurat-Lessard, P.; Sautet, P.; Heyden, A.; Lin, H.; Truhlar, D.G.; Pezeshki, S.; Lin, H.; Riahi, S.; et al. Multiscale Modeling of Chemistry in Water: Are We There Yet? *J. Chem. Theory Comput.* **2013**, *9*, 2231–2241. [CrossRef] [PubMed]
5. Duster, A.W.; Wang, C.H.; Garza, C.M.; Miller, D.E.; Lin, H. Adaptive quantum/molecular mechanics: What have we learned, where are we, and where do we go from here? *WIRES Comput. Mol. Sci.* **2017**, *7*, e1310. [CrossRef]
6. Zhang, Y.J.; Khorshidi, A.; Kastlunger, G.; Peterson, A.A. The potential for machine learning in hybrid QM/MM calculations. *J. Chem. Phys.* **2018**, *148*, 241740. [CrossRef] [PubMed]
7. Steinmann, C.; Reinholdt, P.; Nørby, M.S.; Kongsted, J.; Olsen, J.M.H. Response properties of embedded molecules through the polarizable embedding model. *Int. J. Quantum Chem.* **2018**, *0*, e25717. [CrossRef]

8. Morzan, U.N.; Alonso de Armiño, D.J.; Foglia, N.O.; Ramírez, F.; González Lebrero, M.C.; Scherlis, D.A.; Estrin, D.A. Spectroscopy in Complex Environments from QM–MM Simulations. *Chem. Rev.* **2018**, *118*, 4071–4113. [CrossRef] [PubMed]
9. Dohn, A.O.; Jónsson, E.O.; Kjær, K.S.; van Driel, T.B.; Nielsen, M.M.; Jacobsen, K.W.; Henriksen, N.E.; Møller, K.B. Direct Dynamics Studies of a Binuclear Metal Complex in Solution: The Interplay Between Vibrational Relaxation, Coherence, and Solvent Effects. *J. Phys. Chem. Lett.* **2014**, *5*, 2414–2418. [CrossRef] [PubMed]
10. van Driel, T.B.; Kjær, K.S.; Hartsock, R.W.; Dohn, A.O.; Harlang, T.; Chollet, M.; Christensen, M.; Gawelda, W.; Henriksen, N.E.; Kim, J.G.; et al. Atomistic characterization of the active-site solvation dynamics of a model photocatalyst. *Nat. Commun.* **2016**, *7*, 13678. [CrossRef] [PubMed]
11. Dohn, A.O.; Jónsson, E.Ö.; Levi, G.; Mortensen, J.J.; Lopez-Acevedo, O.; Thygesen, K.S.; Jacobsen, K.W.; Ulstrup, J.; Henriksen, N.E.; Møller, K.B.; et al. Grid-Based Projector Augmented Wave (GPAW) Implementation of Quantum Mechanics/Molecular Mechanics (QM/MM) Electrostatic Embedding and Application to a Solvated Diplatinum Complex. *J. Chem. Theory Comput.* **2017**, *13*, 6010–6022. [CrossRef] [PubMed]
12. Erba, A.; Baima, J.; Bush, I.; Orlando, R.; Dovesi, R. Large-Scale Condensed Matter DFT Simulations: Performance and Capabilities of the CRYSTAL Code. *J. Chem. Theory Comput.* **2017**, *13*, 5019–5027. [CrossRef] [PubMed]
13. Dovesi, R.; Saunders, V.R.; Roetti, C.; Olando, R.; Zicovich-Wilson, C.M.; Pascale, F.; Civalleri, B.; Doll, K.; Harrison, N.M.; Bush, I.J.; et al. *CRYSTAL17 User's Manual*; University of Torino: Torino, Italy, 2017.
14. Dovesi, R.; Erba, A.; Orlando, R.; Zicovich-Wilson, C.M.; Civalleri, B.; Maschio, L.; Rérat, M.; Casassa, S.; Baima, J.; Salustro, S.; et al. Quantum-mechanical condensed matter simulations with CRYSTAL. *Wiley Interdiscip. Rev. Comput. Mol. Sci.* **2018**, *8*, e1360. [CrossRef]
15. Labat, F.; Baranek, P.; Domain, C.; Minot, C.; Adamo, C. Density functional theory analysis of the structural and electronic properties of $TiO_2$ rutile and anatase polytypes: Performances of different exchange-correlation functionals. *J. Chem. Phys.* **2007**, *126*, 154703. [CrossRef] [PubMed]
16. Bai, Y.; Mora-Seró, I.; De Angelis, F.; Bisquert, J.; Wang, P. Titanium Dioxide Nanomaterials for Photovoltaic Applications. *Chem. Rev.* **2014**, *114*, 10095–10130. [CrossRef] [PubMed]
17. Ma, Y.; Wang, X.; Jia, Y.; Chen, X.; Han, H.; Li, C. Titanium Dioxide-based Nanomaterials for Photocatalytic Fuel Generations. *Chem. Rev.* **2014**, *114*, 9987–10043. [CrossRef] [PubMed]
18. Schneider, J.; Matsuoka, M.; Takeuchi, M.; Zhang, J.; Horiuchi, Y.; Anpo, M.; Bahnemann, D.W. Understanding $TiO_2$ Photocatalysis: Mechanisms and Materials. *Chem. Rev.* **2014**, *114*, 9919–9986. [CrossRef] [PubMed]
19. Rajh, T.; Dimitrijevic, N.M.; Bissonnette, M.; Koritarov, T.; Konda, V. Understanding $TiO_2$ Photocatalysis: Mechanisms and Materials. *Chem. Rev.* **2014**, *114*, 10177–10216. [CrossRef] [PubMed]
20. Diebold, U. Perspective: A Controversial Benchmark System for Water-oxide Interfaces: $H_2O/TiO_2$(110). *J. Chem. Phys.* **2017**, *147*, 040901. [CrossRef] [PubMed]
21. Mu, R.; Zhao, Z.j.; Dohnálek, Z.; Gong, J. Structural Motifs of Water on Metal Oxide Surfaces. *Chem. Soc. Rev.* **2017**, *46*, 1785–1806. [CrossRef] [PubMed]
22. De Angelis, F.; Di Valentin, C.; Fantacci, S.; Vittadini, A.; Selloni, A. Theoretical Studies on Anatase and Less Common $TiO_2$ Phases: Bulk, Surfaces, and Nanomaterials. *Chem. Rev.* **2014**, *114*, 9708–9753. [CrossRef] [PubMed]
23. Fazio, G.; Ferrighi, L.; Di Valentin, C. Spherical versus Faceted Anatase $TiO_2$ Nanoparticles: A Model Study of Structural and Electronic Properties. *J. Phys. Chem. C* **2015**, *119*, 20735–20746. [CrossRef]
24. Selli, D.; Fazio, G.; Di Valentin, C. Modelling Realistic $TiO_2$ Nanospheres: A Benchmark Study of SCC-DFTB against DFT. *J. Chem. Phys.* **2017**, *147*, 164701. [CrossRef] [PubMed]
25. Selli, D.; Fazio, G.; Di Valentin, C. Using Density Functional Theory to Model Realistic $TiO_2$ Nanoparticles, Their Photoactivation and Interaction with Water. *Catalysts* **2017**, *7*, 357. [CrossRef]
26. Shirai, K.; Fazio, G.; Sugimoto, T.; Selli, D.; Ferraro, L.; Watanabe, K.; Haruta, M.; Ohtani, B.; Kurata, H.; Di Valentin, C.; et al. Water-Assisted Hole Trapping at Highly Curved Surface of Nano-$TiO_2$ Photocatalyst. *J. Am. Chem. Soc.* **2018**, *140*, 1415–1422. [CrossRef] [PubMed]

27. Li, G.; Li, L.; Boerio-Goates, J.; Woodfield, B.F. High Purity Anatase TiO$_2$ Nanocrystals: Near Room-Temperature Synthesis, Grain Growth Kinetics, and Surface Hydration Chemistry. *J. Am. Chem. Soc.* **2005**, *127*, 8659–8666. [CrossRef] [PubMed]
28. Fazio, G.; Selli, D.; Seifert, G.; Di Valentin, C. Curved TiO$_2$ Nanoparticles in Water: Short (Chemical) and Long (Physical) Range Interfacial Effects. *ACS Appl. Mater. Interfaces* **2018**, *10*, 29943–29953. [CrossRef] [PubMed]
29. Bahn, S.R.; Jacobsen, K.W. An object-oriented scripting interface to a legacy electronic structure code. *Comput. Sci. Eng.* **2002**, *4*, 55–66. [CrossRef]
30. Larsen, A.; Mortensen, J.; Blomqvist, J.; Castelli, I.; Christensen, R.; Dulak, M.; Friis, J.; Groves, M.; Hammer, B.; Hargus, C.; et al. The Atomic Simulation Environment—A Python library for working with atoms. *J. Phys. Condens. Matter* **2017**, *29*, 273002. [CrossRef] [PubMed]
31. Hunt, D.; Sanchez, V.M.; Scherlis, D.A. A quantum-mechanics molecular-mechanics scheme for extended systems. *J. Phys. Condens. Matter* **2016**, *28*, 335201. [CrossRef] [PubMed]
32. Laio, A.; VandeVondele, J.; Rothlisberger, U. A Hamiltonian Electrostatic Coupling Scheme for Hybrid Car–Parrinello Molecular Dynamics Simulations. *J. Chem. Phys.* **2002**, *116*, 6941–6947. [CrossRef]
33. Freindorf, M.; Shao, Y.; Furlani, T.R.; Kong, J. Lennard–Jones parameters for the combined QM/MM method using the B3LYP/6-31G*/AMBER potential. *J. Comput. Chem.* **2005**, *26*, 1270–1278. [CrossRef] [PubMed]
34. Gillan, M.J.; Alfè, D.; Michaelides, A. Perspective: How good is DFT for water? *J. Chem. Phys.* **2016**, *144*, 130901. [CrossRef] [PubMed]
35. Head-Gordon, T.; Hura, G. Water Structure from Scattering Experiments and Simulation. *Chem. Rev.* **2002**, *102*, 2651–2670. [CrossRef] [PubMed]
36. Todorova, T.; Seitsonen, A.P.; Hutter, J.; Kuo, I.W.; Mundy, C.J. Molecular Dynamics Simulation of Liquid Water: Hybrid Density Functionals. *J. Phys. Chem. B* **2006**, *110*, 3685–3691. [CrossRef] [PubMed]
37. Seitsonen, A.P.; Bryk, T. Melting temperature of water: DFT-based molecular dynamics simulations with D3 dispersion correction. *Phys. Rev. B* **2016**, *94*, 184111. [CrossRef]
38. Horn, H.W.; Swope, W.C.; Pitera, J.W.; Madura, J.D.; Dick, T.J.; Hura, G.L.; Head-Gordon, T. Development of an Improved Four-Site Water Model for Biomolecular Simulations: TIP4P-Ew. *J. Chem. Phys.* **2004**, *120*, 9665–9678. [CrossRef] [PubMed]
39. Wikfeldt, K.T.; Batista, E.R.; Vila, F.D.; Jónsson, H. A transferable H$_2$O interaction potential based on a single center multipole expansion: SCME. *Phys. Chem. Chem. Phys.* **2013**, *15*, 16542–16556. [CrossRef] [PubMed]
40. Medders, G.R.; Babin, V.; Paesani, F. Development of a "First-Principles" Water Potential with Flexible Monomers. III. Liquid Phase Properties. *J. Chem. Theory Comput.* **2014**, *10*, 2906–2910. [CrossRef] [PubMed]
41. Cisneros, G.A.; Wikfeldt, K.T.; Ojamäe, L.; Lu, J.; Xu, Y.; Torabifard, H.; Bartók, A.P.; Csányi, G.; Molinero, V.; Paesani, F. Modeling Molecular Interactions in Water: From Pairwise to Many-Body Potential Energy Functions. *Chem. Rev.* **2016**, *116*, 7501–7528. [CrossRef] [PubMed]
42. Babin, V.; Leforestier, C.; Paesani, F. Development of a "First Principles" Water Potential with Flexible Monomers: Dimer Potential Energy Surface, VRT Spectrum, and Second Virial Coefficient. *J. Chem. Theory Comput.* **2013**, *9*, 5395–5403. [CrossRef] [PubMed]
43. Jurečka, P.; Šponer, J.; Černý, J.; Hobza, P. Benchmark database of accurate (MP2 and CCSD(T) complete basis set limit) interaction energies of small model complexes, DNA base pairs, and amino acid pairs. *Phys. Chem. Chem. Phys.* **2006**, *8*, 1985–1993. [CrossRef] [PubMed]
44. Rajh, T.; Chen, L.X.; Lukas, K.; Liu, T.; Thurnauer, M.C.; Tiede, D.M. Surface Restructuring of Nanoparticles: An Efficient Route for Ligand-Metal Oxide Crosstalk. *J. Phys. Chem. B* **2002**, *106*, 10543–10552. [CrossRef]
45. Becke, A.D. Density-functional thermochemistry. III. The role of exact exchange. *J. Chem. Phys.* **1993**, *98*, 5648–5652. [CrossRef]
46. Lee, C.; Yang, W.; Parr, R.G. Development of the Colle-Salvetti correlation-energy formula into a functional of the electron density. *Phys. Rev. B* **1988**, *37*, 785–789. [CrossRef]
47. Gatti, C.; Saunders, V.; Roetti, C. Crystal-field effects on the topological properties of the electron-density in molecular-crystals. The case of urea. *J. Chem. Phys. B* **1994**, *101*, 10686–10696. [CrossRef]
48. Jorgensen, W.L.; Chandrasekhar, J.; Madura, J.D.; Impey, R.W.; Klein, M.L. Comparison of simple potential functions for simulating liquid water. *J. Chem. Phys.* **1983**, *79*, 926–935. [CrossRef]

49. Bates, D.M.; Tschumper, G.S. CCSD(T) Complete Basis Set Limit Relative Energies for Low-Lying Water Hexamer Structures. *J. Phys. Chem. A* **2009**, *113*, 3555–3559. [CrossRef] [PubMed]
50. Temelso, B.; Archer, K.A.; Shields, G.C. Benchmark Structures and Binding Energies of Small Water Clusters with Anharmonicity Corrections. *J. Phys. Chem. A* **2011**, *115*, 12034–12046. [CrossRef] [PubMed]
51. Humphrey, W.; Dalke, A.; Schulten, K. VMD—Visual Molecular Dynamics. *J. Mol. Graph.* **1996**, *14*, 33–38. [CrossRef]
52. Martínez, L.; Andrade, R.; Birgin, E.G.; Martinez, J.M. PACKMOL: A Package for Building Initial Configurations for Molecular Dynamics Simulations. *J. Comput. Chem.* **2009**, *30*, 2157–2164. [CrossRef] [PubMed]
53. Perdew, J.P.; Burke, K.; Ernzerhof, M. Generalized Gradient Approximation Made Simple. *Phys. Rev. Lett.* **1996**, *77*, 3865–3868. [CrossRef] [PubMed]
54. Walser, R.; Mark, A.E.; van Gunsteren, W.F. On the Temperature and Pressure Dependence of a Range of Properties of a Type of Water Model Commonly Used in High-Temperature Protein Unfolding Simulations. *Biophys. J.* **2000**, *78*, 2752–2760. [CrossRef] [PubMed]
55. Schwegler, E.; Grossman, J.C.; Gygi, F.; Galli, G. Towards an assessment of the accuracy of density functional theory for first principles simulations of water. II. *J. Chem. Phys.* **2004**, *121*, 5400–5409. [CrossRef]
56. Grossman, J.C.; Schwegler, E.; Draeger, E.W.; Gygi, F.; Galli, G. Towards an assessment of the accuracy of density functional theory for first principles simulations of water. *J. Chem. Phys.* **2004**, *120*, 300–311. [CrossRef] [PubMed]

**Sample Availability:** Samples of the compounds are not available from the authors.

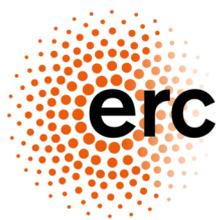
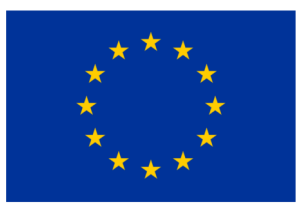